# Weak de-localization in graphene on a ferromagnetic insulating film

*Luca Pietrobon, Lorenzo Fallarino, Andreas Berger, Andrey Chuvilin, Fèlix Casanova, Luis E. Hueso\**

L. Pietrobon, L. Fallarino, A. Berger, A. Chuvilin, F. Casanova, L. E. Hueso
CIC nanoGUNE, 20018 Donostia-San Sebastian, Basque Country, Spain
E-mail: l.hueso@nanogune.eu

Graphene has been predicted to develop a magnetic moment by proximity effect when placed on a ferromagnetic film, a promise that could open exciting possibilities in the fields of spintronics and magnetic data recording. In this work, we study in detail the interplay between the magnetoresistance of graphene and the magnetization of an underlying ferromagnetic insulating film. A clear correlation between both magnitudes is observed but we find, through a careful modelling of the magnetization and the weak localization measurements, that such correspondence can be explained by the effects of the magnetic stray fields arising from the ferromagnetic insulator. Our results emphasize the complexity arising at the interface between magnetic and two-dimensional materials.

## 1. Introduction

Placing in contact a ferromagnetic (FM) and a non-magnetic (NM) material unfolds a plethora of emergent phenomena that have been under intense investigation for several decades in such diverse fields as superconductivity,[1,2] strongly correlated materials[3] and spintronics.[4,5] A classical idea exploiting the FM-NM interface is the magnetic proximity effect (MPE),[6,7] the induced magnetism in a NM material by simple physical proximity to a magnetic one. Graphene is a particularly interesting choice being simultaneously a non-magnetic material and one that can be functionalized. Given its 2D structure, it is the natural probe for an interface effect, like the proximity one, where chemistry does not necessarily play a major role.[8,9] In addition, recent publications predict a large spin polarization of graphene's π orbitals when the carbon lattice is placed on a ferromagnetic insulator.[10–12] An induced magnetic response in graphene could lead to significant advances in topics such as spin transport, spin transfer torque or magnetic random access memories.[13–15] Moreover, it could move graphene closer to the field of topological insulators, by creating one of the building blocks towards the realization of the quantized Anomalous Hall effect in graphene.[16,17,18] As our choice of ferromagnetic material, we selected thick films of Yttrium Iron Garnet (YIG or $Y_3Fe_5O_{12}$). Being a ferromagnetic insulator, YIG is an ideal candidate to induce magnetic surface states without drastically perturbing the electronic conduction. Indeed, it has attracted recent interest in a variety of studies in the field of spintronics, including magnonics,[19] Spin Seebeck effect,[20,21] spin pumping,[22,23] Spin Hall magnetoresistance[24,25] and magnetic gating.[26]

In this work we investigate the transport properties of graphene transferred on YIG. Specifically, we make use of extensive magnetoresistance (MR) measurements to show how the magnetization of the ferromagnetic insulating film influences the resistivity of the graphene strip and, conversely, to what extent such measurements are revealing of the YIG's magnetostatics.

Our device is sketched in **Figure 1**. Single layer graphene, produced by Chemical Vapor Deposition (CVD),[27] is transferred on a YIG film and shaped into a Hall bar geometry (20 μm x 5 μm) by means of Electron Beam Lithography and Reactive Ion Etching in a Ar/$O_2$ atmosphere. A second lithographic step defines the Ti/Au electrodes. Electrical characterization has been carried out with a standard four-point-measurement technique and a 100 μA current amplitude. In our experience a current density of 20 μA/μm ensured minimum contributions from electrical noise, albeit if perhaps slightly heating the graphene by a negligible amount. Typical sample characterization done via Hall effect shows that the graphene is moderately doped (carrier density ≈ 4 × $10^{12}$ $cm^{-2}$) and presents an electrical mobility of 2800 $cm^2$/Vs at a temperature of 10 K, which is in accordance with the typical values we routinely measure for CVD graphene when transferred on $SiO_2$. The YIG sample employed here is a 2 μm thick single crystal film (lateral dimensions: 5mm x 5mm) grown by liquid phase epitaxy on a paramagnetic gallium gadolinium garnet (GGG) (111) substrate at Innovent e. V. (Jena, Germany).

## 2. Magnetoresistance measurements

In order to probe the magnetic coupling between the YIG film and the graphene we perform magnetoresistance measurements sweeping an external magnetic field **H**$_{ext}$ both in intensity and direction. The in-plane MR measurements are of special interest in a 2D system. Both the Lorentz force and the weak localization physics depend only on the out-of-plane component of the magnetic field, and thus they do not contribute to the in-plane field MR.[28] Thus, any MR signal measured in this configuration should come from a coupling of the resistivity of graphene to the magnetization of the YIG (as opposed to a coupling to $H_{ext}$). A potential problem might arise since the graphene is never absolutely flat and wrinkles would inevitably give rise to a finite size area where the flux of the external field is non-zero. As a crosscheck experiment, we always performed analogous measurements for graphene transferred on a $SiO_2$ substrate, which proved that in-plane MR - if any - is beyond our experimental accuracy in this case.

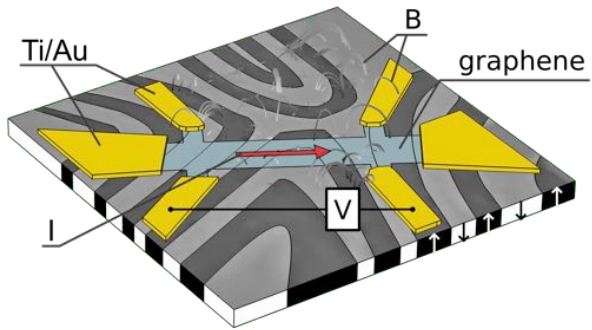

**Figure 1.** Schematic drawing of the device. The YIG sample is represented in its stripe magnetic domain configuration (black and white domains represent the direction of the out-of-plane magnetization). A CVD graphene Hall bar lays over it, contacted through Ti/Au leads. The direction of the current I is highlighted by a red arrow.

**Figure 2**a shows the sheet resistance (ρ) of the graphene device for an in-plane magnetic field. There is a clear non-monotonic modulation of ρ with increasing magnetic field, which is peculiar on its own. In addition, as the in-plane angle φ between the direction of the current in the graphene and the external magnetic field changes (Figure 2b), a six-fold modulation of ρ emerges. The combined dependence of ρ on φ and $H_{ext}$ is shown in Figure 2c-e (ρ in the color scale). Although for temperatures equal or above 50 K ρ shows little dependence on $H_{ext}$ or φ, at low temperatures the sheet resistance depends on both $H_{ext}$ and φ. We can distinguish three regimes. The low-field regime (namely $|H_{ext}| < 50$ Oe), where no angular dependence of ρ is present while its value increases with $H_{ext}$. The second regime ($50 < |H_{ext}| < 150$ Oe), where $\rho(H_{ext}, \varphi)$ is non-isotropic. Finally a third regime ($150$ Oe $< |H_{ext}|$) where the resistivity is, again, isotropic.

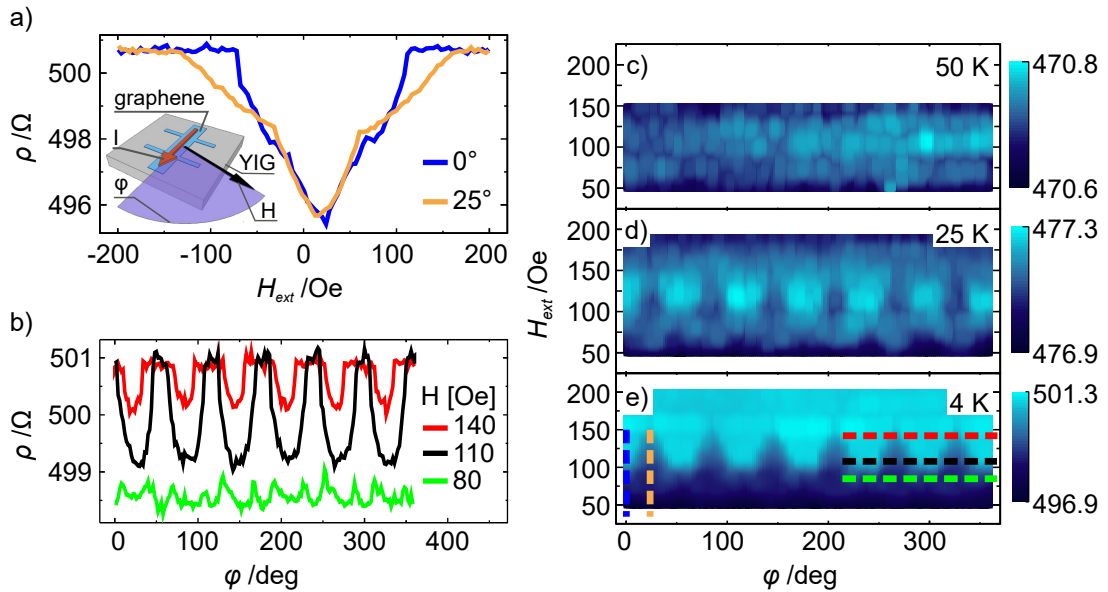

**Figure 2.** Dependence of the sheet resistance ρ of the graphene on the magnitude and orientation of an in plane external magnetic field $\mathbf{H_{ext}}$. a) ρ as a function of $H_{ext}$ at two selected values of the in-plane angle φ, 0° and 25°, at 4 K. We notice both the qualitatively unusual shape of the curve and its angular dependence for 50 Oe < $|H_{ext}|$ <

150 Oe. Inset: schematic representation of the geometry of the experiment. b) $\rho$ as a function of $\varphi$ at 4 K. Three particular values of $H_{ext}$ are shown, namely 80, 110 and 140 Oe for the green, black and red curves respectively. A peculiar 6-fold periodicity is observed for the second graph, which is then lost at higher fields. c, d, e) $\rho$ (color scale) as a function of both intensity and direction of the external, in-plane, magnetic field at 3 different temperatures. Panels a) and b) are cuts to the data in panel e), the dotted lines in the latter being guides to the eye, color coded as the respective curves in the first panels.

### 3. Magnetic characterization

To confirm that these modulations in fact relate to the substrate magnetization, we turn our focus to the ferromagnetic insulator. In **Figure 3**a we show the room temperature magnetic characterization of the YIG performed with a vibrating sample magnetometer (VSM). Under an in-plane magnetic field of magnitude > 100 Oe, the YIG film is magnetically isotropic, but at small in-plane magnetic fields ($H_{ext}$<50 Oe) the hysteresis loop shows an articulated structure, consequence of a non-trivial magnetization process.

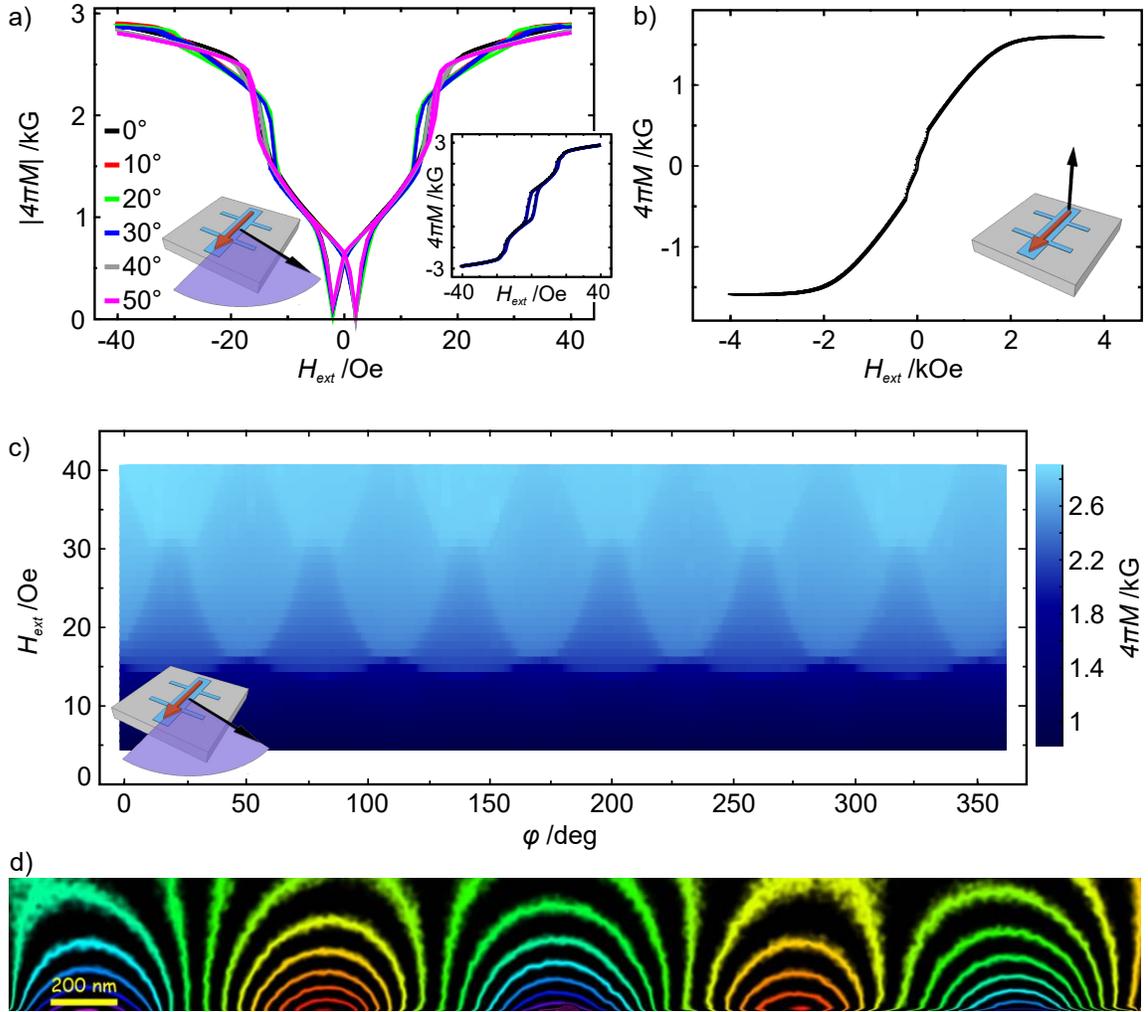

**Figure 3.** Magnetic characterization of the YIG (111) on GGG sample at room temperature, as a function of intensity ($H_{ext}$) and in-plane direction ($\varphi$) of an external magnetic field **$H_{ext}$**. a) Absolute value of the in-plane magnetization M as a function of $H_{ext}$ for selected values of $\varphi$. As **$H_{ext}$** rotates in the plane of the sample, changes in the hysteresis loop are visible for 15 Oe < |$H_{ext}$| < 30 Oe. In the context of this work a positive or negative value of the field **B** = $\mu_0$(**$H_{ext}$**+**M**) at the surface of the YIG contribute equally to the resistance of the device, which makes |M| a more relevant

quantity than M. Inset: original measurements of M from which Figure 3a has been derived. b) Out-of-plane magnetization M of the YIG as a function of an out-of-plane external magnetic field $H_{ext}$. c) $M(H_{ext}, \varphi)$ (color scale) for different values of $H_{ext}$ and $\varphi$. d) Cross-sectional side view image of the stray field over the surface of the YIG obtained by electron holography, the lines on the picture are the field lines of the stray field $H_{str}$ at $H_{ext}=0$. Color represents the direction of the field, i.e. red and blue lines are pointing in opposite directions.

## 4. Discussion

YIG films are well known to exhibit a stripe-domain magnetization pattern.[29,30] The geometry of such domains for films of similar thickness as ours is, in a first approximation, that of parallel, in-plane stripes of opposite magnetization directions. At $H_{ext}=0$ the magnetization of the epitaxial film is directed out of plane, with stray field lines connecting neighbouring domains. To confirm this result in our specific sample as well, we used electron holography on cross-section of the YIG sample. Phase images were reconstructed from the holograms and field lines were built as the isophase lines[30] (see Figure 3d). The reconstruction clearly shows the stray field lines connecting neighbouring domains, confirming the magnetostatic description. Lorentz microscopy images, as the ones elegantly reported by Craus,[29] show that under an external in-plane magnetic field, the average size of the domains grows and the orientation of the magnetization gradually rotates into the plane of the film, until saturation is reached. Specifically, in Figure 3c we note a six-fold symmetry in the measured magnetic moment as the external field rotates in the sample plane. Similarly to what has been observed for the resistivity in Figure 2e, we can again distinguish three regimes in the dependence of the magnetization **M** on **H**$_{ext}$. The first one, occurring at small fields, shows an isotropic magnetization. The second one (for 15 Oe < |H$_{ext}$| < 30 Oe), where **M** depends on φ; and a third one for higher fields, where again **M** is isotropic. YIG has a crystallographic cubic structure which, when observed from the (111) direction, shows an apparent 6-fold symmetry. The crystal symmetry manifests itself as an in-plane magnetization symmetry.[33] Since YIG is a very soft ferromagnet, the presence of such in-plane magnetocrystalline anisotropy is generally overlooked. However, especially in low-field measurements (for instance like the ones reported here or elsewhere in the literature),[34] these details of the YIG magnetic behaviour can play a relevant role in the electronic transport.

There is a strong correlation between the in-plane magnetization data and the ρ(H$_{ext}$) shown in Figure 2e. More quantitatively, the exact values of H$_{ext}$ that delimit the different regimes in Figure 3c are slightly different since the two measurements were carried out at different temperatures (room temperature for the VSM measurements, 4K for the magnetoresistance ones). The critical values of **H**$_{ext}$ for the latter are slightly larger than for the former, which is consistent with a moderate magnetic hardening of the YIG at low temperatures.

Considering the results shown above, it is tempting to speculate that the current **I** in the graphene channel is being spin-polarized via proximity effect to the magnetization **M** of the YIG, and that the product **I**·**M** is at the origin of the modulation of the graphene's resistivity. However, we show here below that a simpler explanation should be considered.

In order to understand the changes in the graphene resistivity with the applied magnetic field, we should consider the well-known quantum phenomenon of weak localization (WL). WL is a positive correction to the resistivity of a conductor, which decreases in an increasing magnetic field. At its core, WL is an interference effect with a natural length scale of the electron's phase coherence length $\lambda_\varphi$,[28] which will be generally material dependent. For the specific case of graphene, we expect a relatively strong WL due to the low dimensionality since a requirement for WL is the presence of self-intersecting sections along the electrons propagating paths, which are more likely in lower dimensions. More in detail, WL arises in graphene as a result of intervalley and intravalley scattering caused by defects.[35] In some cases, weak anti-localization (WAL) has also been observed in graphene samples at low doping and relatively high temperatures.[36] Beyond simple WL, recently large values of magnetoresistance even at room temperature in single layer graphene point out to the complex charge transport in this material due to the presence if inhomogeneous charge distributions.[37] As a general rule, in a flat system only the out-of-plane magnetic field contributes to the changes in resistivity: since the physical quantity at the origin of the de-localization is the flux of the magnetic field **B** through the self-intersecting paths, all the magnetic contributions to the weak localization must be in the out-of-plane direction.

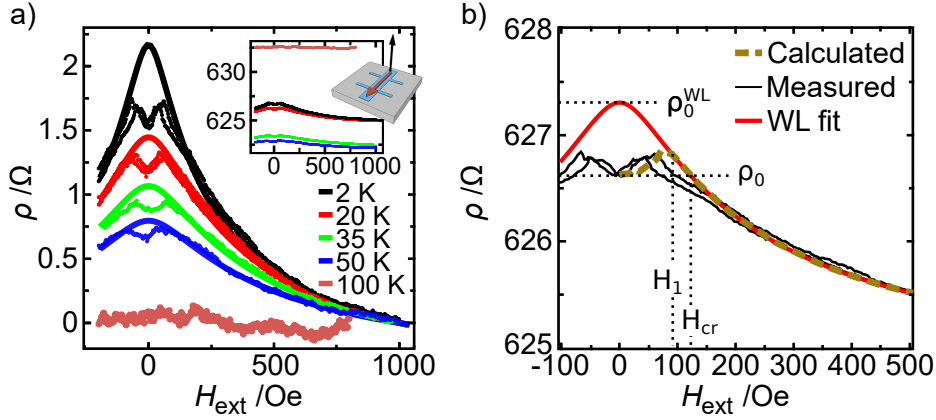

**Figure 4.** a) Measured graphene sheet resistance ρ as a function of an out-of-plane magnetic field $H_{ext}$ at different temperatures, where the $\rho(H_{ext})$ curves have been rigidly shifted for easier comparison (see inset for original data). As the temperature rises from 2 K, both $\rho(H_{ext})$ and the quantity $\rho^{max} - \rho_0$ goes to zero, suggesting that the local minimum in the MR around $H_{ext}=0$ is due to weak-localization physics (the solid lines are fit to the weak localization formula for graphene, see Supplementary Information for the values of the fitting parameters). Sketch: representation of the out-of-plane geometry. b) Closeup of the data in a) for T=2 K, where we add the calculated values for $\langle \rho(H_{ext}) \rangle$ after simulation of a stripe domain structure in the YIG.

In **Figure 4**a we show ρ as a function of an out-of-plane magnetic field together with the fits to the data points for $|H_{ext}| > 200$ Oe (considering the magnetization of YIG as a single domain in this regime) to the WL formula $\rho^{WL}(B)$ for graphene.[35,38,39] For such values of $H_{ext}$, the fit in Figure 4a describes quite accurately the measurements. However, around $H_{ext}=0$ the measurements show a local minimum in ρ, whereas standard WL would predict a (local) maximum.

Taking into account the information outlined above, we argue that the field lines arising from the magnetic domains of the YIG film are the main contributors to the delocalization of the carriers in graphene, causing the minimum at $H_{ext}=0$ in the magnetoresistance. For high values of the applied magnetic field ($|H_{ext}| > 2$ kOe), the magnetization **M** of the YIG is saturated and parallel to the external field (see Figure 3a). As $H_{ext}$ decreases, **M** gradually turns out of plane and the characteristic stripe domain pattern emerges, causing some stray field lines to cross the graphene (Figure 1). In this regime, the flux of the magnetic field **B** through the graphene surface changes sign on a scale comparable to the lateral dimensions $d$ of the magnetic domains, which is expected to be comparable to the film thickness (in our particular case, the YIG film is 2 μm thick). On the other hand, the phase coherence length $\lambda_\varphi$ in graphene - which can be extracted from WL theory – is 10 to 100 times smaller than $d$. We can then expect WL effects in the MR measurements since, even though the flux of B through the whole graphene averages zero, locally (*i.e.* on the scale of $\lambda_\varphi$) WL always gives a positive contribution to the resistivity, just as if the graphene were in a homogeneous magnetic field.

## 5. Quantitative analysis

We can move into more quantitative terms now for a proper fitting of the WL data in the full range of the magnetic field. The fits in Figure 4a effectively consider the YIG as monodomain for $|H_{ext}|>200$ Oe. The WL is a function of the magnetic field $B = \mu_0(H_{ext} + H_{str})$, where here on we distinguish between the external magnetic field $H_{ext}$ and the stray field from the YIG domains $H_{str}$. As long as $H_{str} \neq 0$ (*i.e.*, the YIG is in a multidomain configuration) a fit of $\rho(H_{ext})$ such as in Figure 4a is not rigorously valid. *A priori*, we cannot exclude the presence of stripe domains until the ferromagnet is saturated, which happens only for a field $H_s \sim 2$ kOe. For $H_{ext} < H_s$ the magnetic domains will change in two ways:[30] (i) by *displacement* of domain walls, leading to the growth of the domains in which the magnetization makes an acute angle with $H_{ext}$; (ii) by *rotation* of the magnetization vectors within domains.

The displacement processes tend to dominate at small fields and are well described in the existing literature. Models such as the one presented by Draaisma[40] enable us to evaluate $H_{str}$ as a function of $H_{ext}$. We consider one layer of alternating magnetic domains and compute the stray field $H_{str}(x,z)$, where the (x,y)-plane is taken to be the plane of the YIG surface. As $H_{ext}$ changes, the size $d$ of the domains evolves and a new field $H_{str}(x,z)$ is found. We can then evaluate the resistivity of the graphene at each point in space and average over the length of the device:

$$\langle \rho(H_{ext}) \rangle = \frac{1}{L} \int_0^L dx \, \rho^{WL}(B_{tot}(x))$$

where $\mathbf{B}_{tot}(x) = \mu_0 (\mathbf{H}_{ext} + \mathbf{H}_{str}(x, z_0))$ and $z_0$ is the height of the graphene above the YIG surface. This approach describes the whole magnetization process as a wall-displacement one (see supplementary information). A better agreement can be obtained considering a 2-step magnetization process, where for $|H_{ext}| < H_1$ we approximate the development as purely displacement-driven and for $|H_{ext}| \geq H_1$ we take it to be rotation-driven, with **M** rotating towards the out-of-plane direction. In Figure 4b we show the result of these calculations, which accurately describe the data for $H_1 = 90$ Oe. In addition, the model yields a value for the size of the magnetic domains at $H_{ext}=0$ of $d \approx 0.5$ μm, in accordance with the electron holography images in Figure 3d. The agreement between the experimental data and the model could be improved with a more complex description considering both the displacement and rotation processes at each value of $H_{ext}$, but such level of detail is beyond the scope of this work. Our interest lies in understanding the nature of the coupling between $\rho$ and the magnetic underlayer, which we find can be explained through weak localization physics.

To further support our previous argument, we show in Figure 4a $\rho(H_{ext})$ for different temperatures. We consider the difference between the measured sheet resistance ($\rho_0$) and the extrapolated value from the WL-fit ($\rho_0^{WL}$) at $H_{ext}=0$ as indicative of how effectively the stray fields de-localize the carriers in graphene. An accurate evaluation of quantity $\rho_0^{WL} - \rho_0$ is non trivial, as the fitting process necessarily lacks its most critical parameter, namely the value of $\rho$ when no magnetic field is present (*i.e.* B = 0, instead of simply $H_{ext}=0$). However, a qualitative trend clearly emerges from Figure 4a, where the size of the local minimum at $H_{ext}=0$ (*i.e.* $\rho^{max} - \rho_0$, where $\rho^{max}$ is the maximum value of $\rho(H_{ext})$) decreases with rising temperature. In fact, the ratio $\Delta\rho_{YIG} = [\rho^{max} - \rho_0]/[\rho^{max} - \rho(H_{ext} = 1 \text{ kOe})]$ is approximately 9% for all temperatures where we measure weak localization. Such consistency of $\Delta\rho_{YIG}$ across the measured temperature range further suggests that WL is the effect coupling the magnetization with the changes in $\rho$ in our experiments. Ultimately, the minimum in $\rho$ at $H_{ext}=0$ could qualitatively resemble a weak anti-localization (WAL) phenomenon but this effect for the carrier density in our samples ($4\times10^{12}$ cm$^{-2}$), and it might only present at much lower doping (at best, $2\times10^{12}$ cm$^{-2}$).[35]

Interestingly, the de-localization picture we have presented enables us to reverse the thought process. Instead of calculating the stray field from the YIG and then fitting the WL measurement, we can use the graphene as a sensor for estimating the field at the YIG surface for $H_{ext}=0$. In other words, we can invert the WL formula and indirectly measure the average B field at the surface of the YIG when no external field is applied. Looking at the data in Figure 4a (for clarity, let us consider the one at 2 K), we find that there is a value $H_{cr}>0$ of the external magnetic field such that $\rho^{WL}(\mu_0 H_{cr}) = \rho_0$. Because the function $\rho^{WL}(B)$ is monotonic, we can infer that the flux of B through the graphene is the same at $H_{ext}=0$ and $H_{ext}=H_{cr}$, and conclude that

$$\langle|\mathbf{B}_\perp^{surf}(H_{ext}=0)|\rangle = \mu_0 H_{cr} = [\rho^{WL}]^{-1}(\rho_0)$$

where $\langle|\mathbf{B}_\perp^{surf}|\rangle$ is the component of the **B** field perpendicular to the graphene plane at the surface of the magnetic material, averaged over the area of the graphene strip, and $[\rho^{WL}]^{-1}$ is the inverse of the WL function. The value we extract for this dataset is $H_{cr} \approx$ 125 Oe. Interestingly, we extract the same value from the simulation introduced earlier, with the two-step magnetization approximation. This is fairly consistent with the electron holography measurements in Figure 3d, where the measured average vertical component of $H_{str}$ is of the same order of magnitude (75 Oe).

This measurement of $H_{cr}$ is clearly different from the average magnetic field a magnetometer would measure far from the surface of the sample, as the stray field is stronger the closer to the YIG's surface. Our measurement technique is also different from a micro-SQUID one, as this one would be sensitive to $\langle\mathbf{B}_\perp^{surf}\rangle$ instead of $\langle|\mathbf{B}_\perp^{surf}|\rangle$, the former tending to a value much smaller than the latter as the area of the SQUID increases. A micro-SQUID can deliver very local information about $\mathbf{B}_\perp^{surf}$ (for instance, if the dimensions over which the flux of B is sampled are smaller than the lateral sizes of the stripe domains), and similarly can a Magnetic Force Microscopy measurement. In our case, with MR measurements we determine the average magnetic field at the surface, which is difficult to experimentally evaluate knowing **M** only from a bulk magnetic characterization. Our approach provides an averaged value of $\langle|\mathbf{B}_\perp^{surf}|\rangle$ over a given area, instead of having to scan large areas over which to average $\mathbf{B}_\perp^{surf}$.

## 6. Conclusion

In conclusion, in this manuscript we have presented magnetoresistance measurements of graphene devices placed over a YIG ferromagnetic insulating film. The MR curves remarkably resemble the magnetization of the YIG, both in the six-fold angular symmetry (for the field-in-plane configuration) and in the atypical structure for small fields (in the field-out-of-plane configuration). Although this resemblance might spark discussions regarding the presence of a magnetic proximity effect in graphene when placed in contact with a ferromagnetic layer, we have developed an alternative, more accurate, explanation. In short terms, the stray field from the stripe magnetic domains present in the YIG, which occurs for small values of $H_{ext}$, has an out-of-plane component that de-localizes the electrons in graphene, causing a change in its sheet resistance. Our conclusion is based on both a simulation of the YIG magnetization process, a good fit of the WL equation to our MR curves and on the temperature dependence of our measurements. The measurements presented here could constitute a characterization tool complementary to that of traditional magnetometry for exploring the out-of-plane magnetic field produced by a ferromagnetic sample at its very surface. Finally, we stress how, while exploring magnetic-proximity-effect physics on ferromagnetic insulators,[34] one needs to be very aware of the magnetostatics of the substrate and carefully exclude effects from the unavoidable stray fields.

**Acknowledgements**

This work has been supported by the European Union 7th Framework Programme under the Marie Curie Actions (256470-ITAMOSCINOM, 264034 - Q-NET) and the European Research Council (257654-SPINTROS), as well as by the Spanish MINECO under Project No. MAT2012-37638. L. Fallarino thanks the Basque Government for the PhD fellowship (Grants No. PRE_2013_1_974 and No. PRE_2014_2_142). We would like to thank Amilcar Bedoya-Pinto for the VSM measurements and Oihana Txoperena for the AFM data..

# Supporting Information

**Weak de-localization in graphene on a ferromagnetic insulating film**

*Luca Pietrobon, Lorenzo Fallarino, Andreas Berger, Andrey Chuvilin, Fèlix Casanova, Luis E. Hueso\**

## S1. Control experiment: graphene on SiO$_2$

In the main text we report on modulations of the sheet resistance of graphene ρ in relation to the magnetization state of an underlying YIG film. Specifically, a modulation of ρ is measured as a function of an in-plane external magnetic field. In order to confirm that such modulation is related to the YIG substrate we consider, as control experiment, a graphene Hall bar on a Si/SiO2 substrate with a similar roughness to the samples on a YIG substrate (0.4 nm versus 0.5 nm of rms roughness on a 1 micron square area). The results are shown in Figure 1.

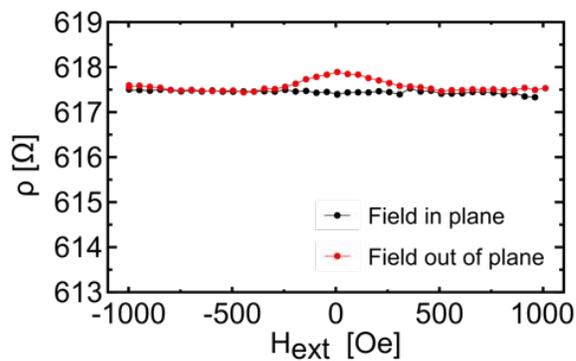

*Figure 1. Control experiment. a) Sheet resistance ρ of a graphene Hall bar on a Si/SiO2 substrate as a function of an external magnetic field $H_{ext}$ at 4 K and 0.1 mA current, both for in-plane and out-of-plane directions. No modulation of ρ is found in the in-plane geometry, in contrast to the case of graphene on YIG shown in Figure 2a in the main text. A direct comparison is not possible due to the different sheet resistances measured, but the modulation of almost 1% for graphene on YIG (Figure 2a) is clearly missing in graphene on Si/SiO$_2$ (in-plane geometry).*

## S2. Determination of the WL parameters

The analytical expression for the resistivity in graphene with the weak-localization corrections ($\rho^{WL}(B)$) was developed by McCann [1]. We report it here for the reader's convenience

$$\rho^{WL}(B) = \rho_0 - \frac{e^2 \rho_0^2}{\pi h}\left[F\left(\frac{B}{B_\phi}\right) - F\left(\frac{B}{B_\phi + 2B_i}\right) - 2F\left(\frac{B}{B_\phi + B_*}\right)\right]$$

$$F(z) = \ln z + \psi\left(\frac{1}{2} + \frac{1}{z}\right), \quad B_{\phi,i,*} = \frac{\hbar}{4e\lambda_{\phi,i,*}^2}$$

where $\psi$ is the digamma function.

The expression for $\rho^{WL}(B)$ comes with 4 independent parameters ($\lambda_\phi, \lambda_i, \lambda_*$ and $\rho_0$), the most relevant of which (especially for small fields) are the value of the resistivity at zero $B$ field ($\rho_0$) and the phase coherence length ($\lambda_\phi$). As pointed out in the main text, $\rho^{WL}$ is a function of the magnetic field $\mathbf{B}=\mu_0(\mathbf{H}_{ext} + \mathbf{H}_{str})$ arising from the superposition of the external field $H_{ext}$ and the stray field of the magnetic substrate $H_{str}$. Typically the last one is unknown except for the case of a magnetically saturated YIG, where $H_{str}=0$. As a first option, we could extract the values of $\rho_0$ and $\lambda_\varphi$ from graphene samples fabricated on $SiO_2$ substrates, but it is our experience that with CVD graphene samples these parameters vary significantly from sample to sample (*e.g.* a 1% change in resistivity across samples is too large of a variation compared to the weak localization corrections of ~ 0.2% reported in the main text). Alternatively, we could fit $\rho^{WL}$ only for $H_{ext}>H_s$ and extrapolate to lower fields, however the saturation field $H_s$ for our sample is 1800 Oe (at room temperature), far larger than the values of interest in this work ($|H_{ext}| \lesssim 500$ Oe). In addition, from our experience on analogous graphene devices on $SiO_2$, we expect relatively small changes in $\rho$ for such high fields, which would make the extrapolation of $\rho_0$ and $\lambda_\varphi$ less accurate.

We propose a different (and, to the best of our knowledge, new) setup for WL measurements, which takes advantage of the 2D nature of the conductor.

Figure 2 shows the measured resistivities $\rho$ as a function of an external fixed $H_{ext}$ at a polar angle $\theta > 0$ with respect to the normal to the YIG surface. The low dimensionality of the graphene makes it sensitive only to the out-of-plane component $B_z$ of the total magnetic field:

$$\begin{aligned} B_z &= \mu_0(\mathbf{H}_{ext} + \mathbf{H}_{str})\cos\theta \\ &= \mu_0(H^z_{ext} + H_{str}(\theta)) \end{aligned}$$

where we write $H_{str}(\theta)$ to emphasize that the stray field will have a non-trivial dependence on $\theta$. This can be seen, for instance, considering the dependence on $\theta$ of the YIG saturation fields (as confirmed by VSM measurements): for $H_{ext}=300$ Oe and $\theta=0$ the sample is *not* saturated ($H_{str} \neq 0$), but for $H_{ext}=300$ Oe and $\theta=90°$ it is ($H_{str}=0$). From this observation we can conclude that $H_{str}$ does indeed depend on θ, although the dependence is more elaborate than $\propto \cos(\theta)$ (otherwise we would assume that even the smallest in-plane magnetic field is sufficient to saturate the sample).

In Figure 2b-c we also show the measured ρ as a function of $H_{ext}^z$: in this case, all the θ-dependent contributions must come from $H_{str}(\theta)$, while the rest of the data must be in a regime where $H_{str}(\theta)$ is negligible with respect to $H_{ext}^z$. A fit to the non-angular-dependent data yields the values for the parameters of $\rho^{WL}(B)$.

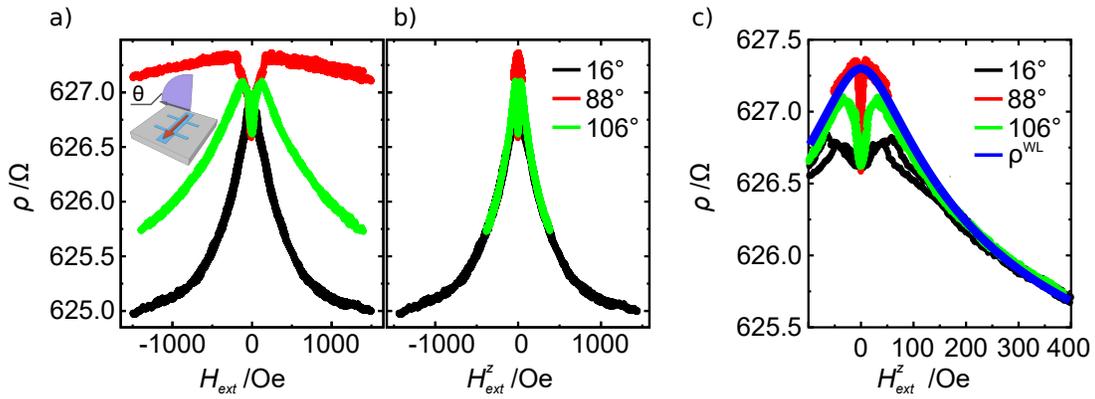

*Figure 2. a) Measured graphene resistivity ρ for an external magnetic field $H_{ext}$ changing both in intensity and direction (polar angle θ>0. b) Same data as in a), but plotted against the out-of-plane component $H_{ext}^z$ of the external field: the angular-dependent data in this plot highlights the regions of field intensity where $H_{str} > H_{ext}^z$. c) Closeup of b).*

In summary, this analysis allows us to distinguish between the contributions to ρ coming from the external field $H_{ext}$ and the ones coming from the YIG stray field $H_{str}$. It also shows that there is a particular field $H_{tr}(\theta)$ above which $H_{str}$ is negligible compared to $H_{ext}$, which takes values up to ≲ 200 Oe. The magnetostatic insight is therefore that, for fields $H_{ext}>H_{tr}$, the difference in magnetic moment among stripe domains are small and the magnetization evolves mainly by rotation.

## S3. Weak Localization parameters

In Figure 4a in the main text we show the Weak Localization fits to the measurements at different temperatures. We report here the fitting parameters; the resistance at zero field, the dephasing length, the intervalley scattering length and the intravalley scattering length, respectively.

| T [K] | $\rho^{WL}(B=0)$ [Ω] | $\lambda_\varphi$ [nm] | $\lambda_i$ [nm] | $\lambda_*$ [nm] |
|---|---|---|---|---|
| 2  | 627.30 ± 0.02 | 223.6 ± 0.3 | 368.6 ± 0.3 | 0.1 ± 0.1 |
| 20 | 626.44 ± 0.02 | 161 ± 2 | 307 ± 2 | 9 ± 2.4 |
| 35 | 623.63 ± 0.01 | 148.8 ± 0.7 | 334 ± 3 | 0.7 ± 0.4 |
| 50 | 623.06 ± 0.02 | 139.3 ± 0.1 | 346.0 ± 0.7 | 3.0 ± 0.5 |

## S4. Magnetization curves for the YIG sample at different temperatures

In the main text we show how the magnetization of the YIG changes with the in-plane direction of the field at room temperature. In a separate study we also verified how the magnetic moment of the YIG sample changes with temperature. Specifically we can see how the saturation field changes from ≈ 40 Oe to 100 Oe.

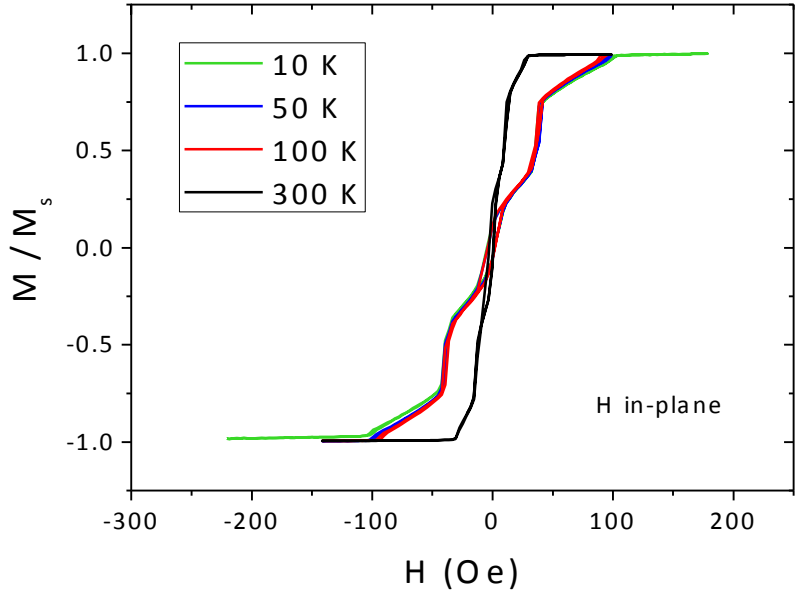

Figure 3. In-plane magnetization of a YIG substrate at different temperatures.

## S5. Modeling the magnetostatics

In order to understand the measurements at small fields we need to calculate the stray field $H_{str}$ coming form the stripe domain magnetization of the substrate. We use the model in [2] and, with reference to the nomenclature used there, our case has N=1, t=1.7μm, and τ=0.2 erg/cm$^2$ [3]. For a given value of magnetization *m*, we compute the domain periodicity *d(m)* and the stray field $H_{str}(m,x,z)$. Equation (6) in [2] enables us to compute also the value of $H_{ext}(m)$ required to reach a magnetization *m*, so that ultimately we obtain the total magnetic field at any point in space as a function of the externally applied field: $H_{tot}(H_{ext}, x,z) = H_{ext} + H_{str}(H_{ext},x,z)$. We can then calculate the resistivity of the graphene at each point and average over the length of the device:

$$\langle \rho(H_{ext}) \rangle = \frac{1}{L} \int_0^L dx\, \rho^{WL}(\mu_0 H_{tot}(x, z_0))$$

where $z_0$ is the effective height of the graphene above the YIG surface. Stripe domains for thick films with perpendicular magnetic anisotropy do not extend all the way to the surface, but have a cap-like structure in the surface region which is primarily magnetized in-plane (and thus by itself not producing significant stray fields). Thus, the effective distance of the graphene from the perpendicular domain structure is substantially larger than the actual surface-to-graphene distance. The main features in the data we want to recover are: (i) the value of $\rho_0 = \rho(H_{ext}=0)$; (ii) an increase of $\rho$ with $H_{ext}$ in the (0, 100) Oe interval; (iii) a local maximum at $H_{ext} \approx 100$ Oe; (iv) WL-like tails for high values of $H_{ext}$.

In Figure 3a we show a selection of calculated curves for different values of the parameter $z_0$. On a qualitative level, we can see the WL tails (iv) and a shoulder for $H_{ext}>0$ (iii), but the quantitative analysis is not quite satisfactory, particularly for the data at small fields.

In light of the analysis of the angular dependence of $\rho^{WL}$, we propose a picture where the changes in $\rho$ are dominated by $H_{str}$ for small fields, whereas above a threshold $H_{ext} = H_1$ the dominant contributor is $H_{ext}$. The ratio $H_1/H_{cr}$ (see Figure 4b in main text) is determinant for the (non)monotonicity of the simulated curves in Figure 3: as shown in Figure 3d, for $H_1<H_{cr}$ the decrease in $H_{str}$ with increasing $H_{ext}$ is rapid enough to yield a non-monotonic behavior of $B^z = \mu_0(H_{ext}^z + H_{str}^z)$, while for $H_1 > H_{cr}$ we have that $B^z$ grows monotonically with $H_{ext}$. As our measurements all show a non-monotonic behavior, we take $H_{cr}$ as an upper limit for $H_1$. Additionally, we note that adapting the model in [2] by introducing an effective distance $z_0$ is effectively like reducing the value of $M_s$, as can be seen by noting that the calculated sheet resistance $\rho_0^{sim} = \rho^{sim}(H_{ext} = 0)$ in Figure 3d (which are for $4\pi\ M_s=300$ G and $z_0 = 10$ nm) are the same as the $\rho_0^{sim}$ calculated in Figure 3b ($4\pi\ M_s=1800$ G, $z0 = 250$ nm).

As it turns out, the data can be well described by this approximation, reproducing each of the (i)-(iv) features we are interested in.

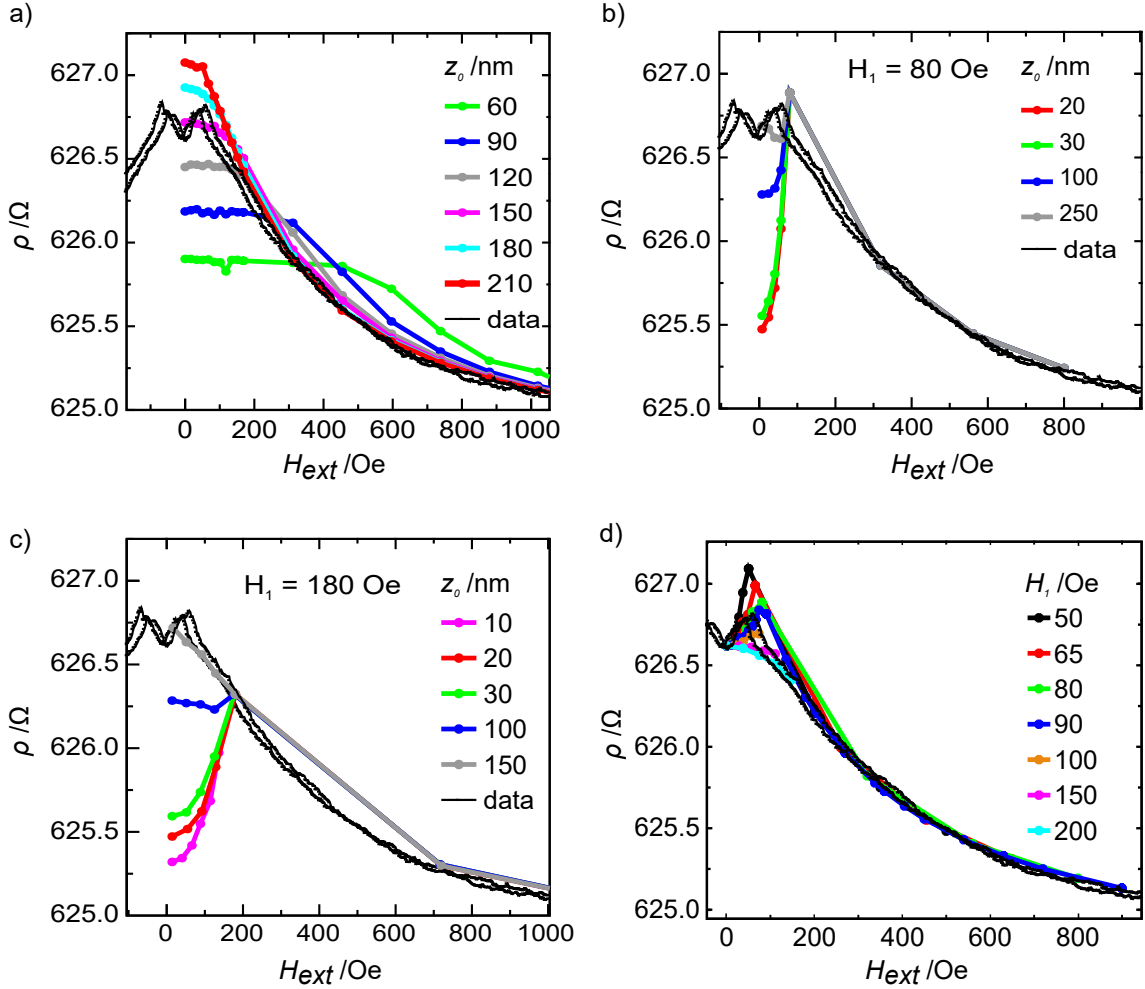

*Figure 4. Calculation of the expected resistivity ⟨ρ⟩ after averaging in space the contributions of $\rho^{WL}(B(x,z_0))$ over the length of the device, where measurements of ρ at 2K (same as in the main text) stand in the background. a) simulations for different heights $z_0$ of the graphene over the YIG surface, based on the model in [2]: the simulation lacks quantitative agreement with the measurements. b) Simulations for different values of $z_0$ after introducing the approximation described in the main text, where we separate between a displacement-driven and a rotation-driven magnetization process at $H_1 = 80$ Oe. c) Same as b), but with $H_1 = 180$ Oe. d) Simulations at $z_0 = 10$ nm*

*and $4\pi M_s = 300$ G, for different values of $H_1$. We can see how the calculated curves change between a monotonic and a non-monotonic behaviour as a function of H1.*